\documentclass{aa}  
\usepackage{graphicx}
\usepackage{txfonts}

\begin{document}

   \title{ALMA resolves molecular clouds in the metal poor Magellanic Bridge A}

   \author{M. T. Valdivia-Mena\inst{1}, M. Rubio\inst{1}, A. D. Bolatto\inst{2}, H. P. Salda\~no\inst{3} \and C. Verdugo\inst{4}
          }

   \institute{Departamento de Astronom\'ia, Universidad de Chile, Santiago, Chile\\ \email{maria.valdivia@ug.uchile.cl}
         \and
             University of Maryland, MD, USA
        \and
             Observatorio Astron\'omico de C\'ordoba, UNC, Argentina
        \and
             Joint Alma Observatory (JAO), Alonso de C\'ordova 3107, Vitacura, Santiago, Chile}

   \date{}
   
   \titlerunning{ALMA Resolves Molecular Clouds in the Metal Poor Magellanic Bridge A}
   \authorrunning{M. T. Valdivia-Mena et al.}

 
  \abstract
   {The Magellanic Bridge is a tidal feature located between both Magellanic Clouds, containing young stars formed in situ. Its proximity allows high-resolution studies of molecular gas, dust and star formation in a tidal, low metallicity environment. }
   {Our goal is to characterize gas and dust emission in Magellanic Bridge A, the 
   source with the highest 870~$\mu$m excess of emission
   found in single dish surveys.}
   {Using the ALMA telescope including the Morita Array, we mapped with sub-parsec resolution a 3\arcmin\, field of view centered on the Magellanic Bridge A molecular cloud, in 1.3 mm continuum emission and \element[][12]{CO}(2$-$1) line emission. This region was also mapped in continuum at 870~$\mu$m and in \element[][12]{CO}(2$-$1) line emission at $\sim6$ pc resolution with the APEX telescope. To study its dust properties, we also use archival Herschel and Spitzer data. We combine the ALMA and APEX \element[][12]{CO}(2$-$1) line cubes to study the molecular gas emission.}
   {Magallanic Bridge A breaks up into two distinct molecular clouds in dust and \element[][12]{CO}(2$-$1) emission, which we call North and South. Dust emission in the North source, according to our best parameters from fitting the far-infrarred fluxes, is $\approx3$ K colder than in the South source in correspondence to its less developed star formation. Both dust sources present large submillimeter excesses in LABOCA data: according to our best fits the excess over the modified blackbody (MBB) fit to the Spitzer/Herschel continuum is $E(870\mu m)\sim7$ and $E(870\mu m)\sim3$ for the North and South sources respectively. Nonetheless, we do not detect the corresponding 1.3 mm continuum with ALMA. Our limits are compatible with the extrapolation of the MBB fits and therefore we cannot independently confirm the excess at this longer wavelength.
   The \element[][12]{CO}(2$-$1) emission is concentrated in two parsec-sized clouds with virial masses around 400 and 700 M$_{\sun}$ each. Their bulk volume densities are $n(H_2)\sim0.7-2.6\times10^3$ cm$^{-3}$, larger than typical bulk densities of Galactic molecular clouds. The \element[][12]{CO} luminosity to H$_2$ mass conversion factor $\alpha_{CO}$ is 6.5 and 15.3 M$_\sun$\,(K\,km\,s$^{-1}$\,pc$^2$)$^{-1}$ for the North and South clouds, calculated using their respective virial masses and \element[][12]{CO}(2$-$1) luminosities. Gas mass estimates from our MBB fits to dust emission yields masses $M\sim1.3\times10^3$ M$_\sun$ and $2.9\times10^3$ M$_\sun$ for North and South respectively, a factor of $\sim4$ larger than the virial masses we infer from \element[][12]{CO}.}
   {}


   \maketitle
%

\section{Introduction\label{sec:intro}}

The Magellanic Bridge was first described through neutral hydrogen (\ion{H}{I}) observations as a gaseous bridge joining the Large Magellanic Cloud (LMC) and Small Magellanic Cloud \citep[SMC,][]{hindman1963}. It is a filamentary structure with lumps, holes and shells, with an extent of 15 to 21 kpc, a total gas mass of $\sim1.5 \times 10^8$ M$_{\sun}$ and an \ion{H}{I} column density between $10^{20}$ and $10^{21}$ cm$^{-2}$ \citep{staveley-smith1998,muller2003a}. Simulations suggest that the Magellanic Bridge is the result of a close gravitational interaction between the LMC and SMC that happened around 200-300 Myrs ago \citep{murai1980, gardiner1994, besla2012}. Star formation has and is taking place in the Magellanic Bridge: there are young stars and clusters less than 100 Myrs old \citep{bica1995, harris2007, bica2015, kalari2018} and evidence of current star formation in the form of H$\alpha$ filamentary shells \citep{muller2007} and young stellar objects \citep[YSO,][]{chen2014}. Its metallicity is low and seems to present a gradient, from values similar to the SMC main body (, \citep[$Z\sim1/5\,Z_{\sun}$,][]{lee2005} to about half (or even less) of the SMC metallicity  \citep[$Z\sim1/10\,Z_{\sun}$,][]{rolleston2003,lehner2008}. At a distance of almost 60 kpc, using the SMC as reference \citep{harries2003,cioni2000}, the Magellanic Bridge is the closest low-metallicity, tidally influenced region outside of a dwarf galaxy, allowing detailed studies of star formation under unique physical conditions.

Stars form in molecular clouds, which are composed of molecular hydrogen (H$_2$) gas and dust. Studying the molecular component of low-metallicity galaxies helps to elucidate how star formation undergoes in metal deficient environments, like those present in high redshift galaxies. To this end, the molecular component of the Magellanic Bridge has been studied through carbon monoxide (\element[][]{CO}) and dust emission, both tracers of H$_2$ gas, but these emissions are weaker than in the SMC main body \citep{meixner2013,muller2014} and, therefore, harder to detect. 

\begin{figure*}[t]
\centering
\includegraphics[width=\hsize]{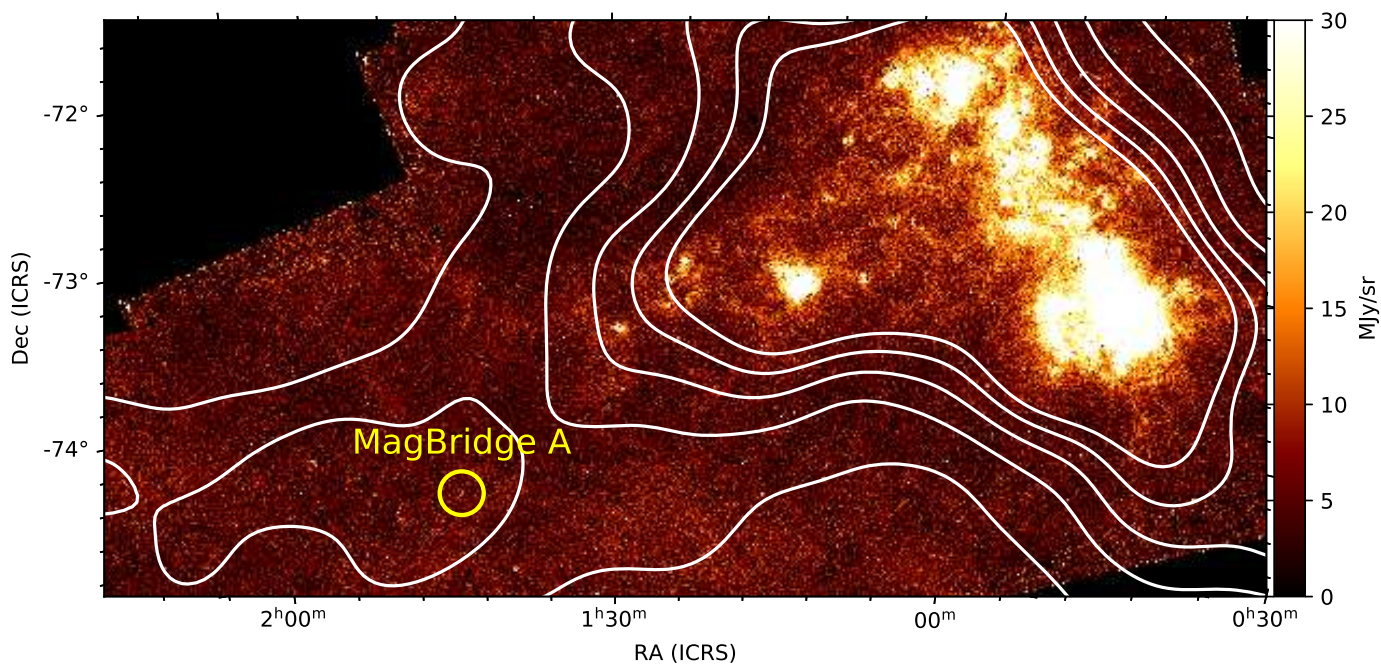}
\caption{\label{fig:SMCreference}Herschel 160 $\mu$m map of the SMC from \citet{meixner2013}. White contours represent the \ion{H}{I} column density map constructed using the Parkes Galactic All-Sky Survey (GASS) first data release \citep{mcclure-griffiths2009}. The contour levels are (1, 1.5, 2, 2.5 and 3 ) $\times 10^{21}$ cm$^{-2}$. The yellow circular region marks the position of Magellanic Bridge A (labeled MagBridge A).}
\end{figure*}

The first study that detected a molecular cloud in the Magellanic Bridge was done by \citet{muller2003b}, using \element[][12]{CO}(1$-$0) observations. The detection was found in a region with an \ion{H}{I} intensity peak and where the 60-to-100 $\mu$m intensity ratio $S_{60\mu m}/S_{100\mu m} < 0.2$. Using a different criteria, matching \ion{H}{I} and 100 $\mu$m intensity peaks, \citet{mizuno2006} detected seven more \element[][12]{CO} clouds which, together with the \citet{muller2003b} detected cloud, have \element[][12]{CO}(1$-$0) intensities between 30 and 140 mK\,km\,s$^{-1}$ in a 2\farcm6 beam.  These molecular clouds have narrow velocity widths ($\Delta v \lesssim 2$ km\,s$^{-1}$), similar to the clouds in the metal-poor far edges of our galaxy, and molecular masses between $(1-7) \times 10^3$ M$_{\sun}$, derived from their CO luminosities \citep{mizuno2006}, similar to the masses of nearby molecular clouds in the Milky Way \citep{mizuno1995, mizuno2001b}. The eight clouds have been labeled from A to H, which is the naming convention that we use in this work. \citet{mizuno2006} concluded that these clouds are a recent product of the Magellanic Bridge ISM and not remnants from the SMC. All of the clouds except for one are within a few tens of parsecs to the closest OB stellar association \citep{bica1995}. YSO candidates have been found towards all of these sources, except for cloud D \citep{chen2014}, showing the correlation of these molecular clouds with local star formation. 

Dust, which is heated by star radiation and emits continuum at the far-infrared (FIR) and millimeter wavelengths, also traces H$_2$ gas. Observations in the FIR and millimeter regime have been done in the part of the Magellanic Bridge that is closest to the SMC, a section usually referred to as the SMC Tail \citep{gordon2011, meixner2013}. These observations reveal that dust in the Magellanic Bridge has a lower brightness than the SMC dust and a gas-to-dust ratio $\sim$1200, similar to the SMC gas-to-dust ratio \citep{gordon2011}. \citet{gordon2009} determined that dust in the \citet{mizuno2006} molecular clouds B and C have temperatures around 16-17 K and a gas-to-dust ratio lower than the rest of the Bridge, between 250-440. 

The focus our work is Magellanic Bridge source A, the molecular cloud that is closest to the SMC ($\sim3.5$ kpc from the main body of the SMC, see Fig. \ref{fig:SMCreference}). Magellanic Bridge A is one of the faintest of the 8 molecular clouds detected by \citet{mizuno2006}, with a \element[][12]{CO}(1$-$0) intensity of 30 mK\,km\,s$^{-1}$ and a molecular mass of $10^3$ M$_{\sun}$,  estimated from its \element[][12]{CO}(1$-$0) emission. 
Magellanic Bridge A is located between two faint, compact \ion{H}{II} regions \citep{muller2007} and is within a stellar association \citep{bica1995}. There is recent star formation in this region, revealed by two YSO candidates within the molecular cloud \citep{chen2014} and its proximity to \ion{H}{II} regions and 24~$\mu$m bright sources \citep{muller2014}. \citet{verdugo2012} found that Magellanic Bridge A has a very high excess emission at 870~$\mu$m in comparison with other dust sources in the LMC and SMC. 

The goal of this work is to characterize gas and dust present in the Magellanic Bridge A molecular cloud at sub-parsec resolution. The Atacama Large Millimeter/Submillimeter Array (ALMA) telescope has the power to resolve gas and dust emission at the desired scales. With this characterization, we can obtain the molecular gas properties in a region which is similar to low-metallicity galaxies at high redshift. Also, we investigate the dust component of Magellanic Bridge A to characterize the submillimeter excess detected in this region \citep{verdugo2012}. To achieve our goal, we use high resolution ALMA observations, together with observations from the Atacama Pathfinder Explorer (APEX) telescope observations, of Magellanic Bridge A, which reveal molecular gas and dust emission.

This article is organized as follows. In \S \ref{observations}, we present new ALMA B6 continuum and \element[][12]{CO}(2$-$1) line observations and complementary APEX observations, together with details of their reduction and imaging. In \S \ref{sec:duststudies}, we describe the submillimeter and millimeter emission 
and characterize the dust emission through its spectral energy distribution.
In \S \ref{molclouds}, we describe the molecular gas emission using \element[][12]{CO}(2$-$1) and determine the physical properties of the detected \element[][12]{CO}(2$-$1) sources. In \S \ref{discussion}, we discuss the results obtained for molecular gas and dust and compare them with previous observations of the ISM in Magellanic Bridge A. Finally, we summarize our results in \S \ref{conclusions}.


\section{Observations and data reduction\label{observations}}

In the following section, we describe the observations used to study the millimeter and submillimeter emission from Magellanic Bridge A. We summarize the main properties of the observations and describe the reduction process for the new data presented in this work.

\subsection{ALMA observations}

Magellanic Bridge A was observed with ALMA, located at Llano de Chajnantor in Atacama, Chile. We use Band 6 observations taken with the 12m array and the 7m array (also known as Morita Array), which include continuum emission observations and the \element[][12]{CO}(2$-$1) line emission data. From now on, we refer to the \element[][12]{CO}(2$-$1) line as CO(2$-$1). Observations were performed in separate cycles: we used 7m array data taken during Cycle 1 and 12m array observations taken in Cycle 2. The datasets belong to project 2012.1.00683.S (PI M. Rubio). As the observations were obtained and reduced in different cycles, we decided to start from the raw data and reprocess completely both datasets, instead of using the delivered data products by ALMA. 

The 12m observations were constructed with 33 antennas and consist of a mosaic of 93 pointings, with 13\farcs7 spacing, and a total excecution time of 107 minutes. Uranus was used as flux calibrator, J0057$-$7040 as phase calibrator and J2357$-$5311 as bandpass calibrator. 

The 7m observations were conducted with 10 antennas and consist of a mosaic of 34 pointings, with a distance of $23\farcs4$ between the center of each pointing, and a total integration time of 93.7 minutes. Uranus was also used as a flux calibrator for these observations. The phase calibrator was J0102$-$7546 and the bandpass calibrator was J2357$-$5311.

We reduce, combine and image the data using the Common Astronomy Software Applications (CASA) package version 4.7, with the consideration that the software was released later than the observations. We use the standard scripts provided by the ALMA Science Archive\footnote{http://almascience.eso.org//aq/} with the delivered raw data to recover the calibrated datasets. The 12m and 7m data are reduced separately. 

We combine the calibrated 12m and 7m observations into one dataset. The observations are done using both the 12m and 7m array to have high resolution data without losing extended emission: the 12m array configuration gives us observations with high resolution, yet it filters out diffuse emission that is larger than its maximum recoverable scale (MRS), which for our 12m observations is 18\farcs3. The MRS increases to 30\farcs9 after the 7m array data are combined with the 12m array observations. To combine data taken with two different arrays of antennas, it is crucial that the relative weights of each visibility are correct, as specified in the CASA guide\footnote{https://casaguides.nrao.edu/index.php/DataWeightsAndCombination}. We calculate the visibility weights of the 12m and 7m observations with the \texttt{statwt} task. After calculating the correct visibility weights, the 12m array and 7m array data were combined using the \texttt{concat} task with a frequency tolerance of 10 MHz (using the \texttt{freqtol} parameter). 

We image the combined 12m and 7m data and generate a continuum emission image and a CO(2$-$1) line cube. We describe the the imaging process for the continuum map in \S \ref{almacont} and the procedure to generate the CO(2$-$1) line cube in \S \ref{almalinecube}. We also combine the ALMA CO(2$-$1) 12m and 7m data with single dish CO(2$-$1) observations obtained with the APEX telescope to include the zero spacing emission. The combination of ALMA and APEX data is described in \S \ref{linecube}.

\subsubsection{1.3 mm continuum\label{almacont}}

\begin{figure}[htb]
   \centering
   \includegraphics[width=0.8\hsize]{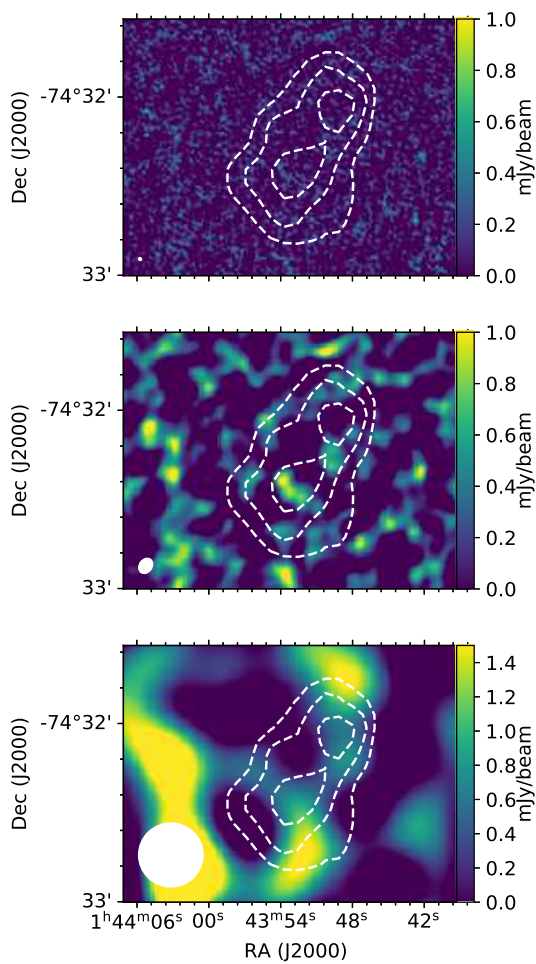}
   \caption{ALMA continuum images at 1.3 mm of Magellanic Bridge A, generated using 12m and 7m data, where imaging is performed using natural weight. The top image corresponds to the continuum without tapering in the visibility data. The middle image is the continuum with its visibility data tapered using a Gaussian with an on-sky FWHM of 5\arcsec. The bottom image is the tapered continuum image convolved to reach a resolution of 22\arcsec. We note that the color scale for this last image is different from the other two images. The white ellipses in the bottom left corner of each image represent the beam sizes. Dashed contours correspond to continuum emission in the APEX 870~$\mu$m map at 25, 30 and 35 mJy/beam.  \label{fig:almacont}}
\end{figure}

We use the standard \texttt{clean} task on the 12m and 7m combined data to generate the 1.3 mm (230 GHz) continuum image, removing the spectral channels that contain the CO(2$-$1) line. We generate the image using both natural weight and briggs weight. The natural weight lowers the noise and allows to detect point sources more effectively, but degrades the angular resolution; on the other hand briggs weighting (also known as robust) provides better resolution and smaller sidelobes of the beam, with a penalty in noise. The briggs image is generated with a robust parameter $r=0.5$. The briggs and natural weighted images resulted in nearly identical rms, 0.18 mJy/beam for briggs weight and 0.17 mJy/beam for natural weight. We decide to use natural weight for this work, as sensitivity tends to be higher for natural weighted images than briggs weighted ones. The natural weighted continuum image is in Figure \ref{fig:almacont}. The continuum image has a field of view (FOV) of 3\arcmin$\times$3\arcmin. 

The continuum image we use for this work is generated using natural weight and a tapering in the u-v plane. The taper gives more weight to the shorter baselines, which might improve the sensitivity to extended sources but degrades the resolution of the image. We apply a tapering of 5\arcsec, which results in an image with a resolution of $5\farcs9 \times 4\farcs7$. To compare with the APEX LABOCA 870~$\mu$m image (see \S \ref{laboca}), we convolved the natural, uv-tapered image to reach a beam FWHM equal to 22\arcsec. The beam size (described through the FWHM in the major and minor axes $\theta_{\text{major}}$ and $\theta_{\text{minor}}$) and rms $\sigma$ of the continuum image, before and after tapering and convolution, are summarized in Table \ref{almacontprops}. The tapered continuum images (convolved and not convolved to 22\arcsec) are in Figure \ref{fig:almacont}.

\begin{table}[ht]
\centering
\caption{Properties of the resulting continuum at 1.3 mm (230 GHz).} 
\label{almacontprops}
\begin{tabular}{lllll}
\hline\hline
Weight    & $\theta_{\text{major}}$  & $\theta_{\text{minor}}$ & PA  & $\sigma$ \\
 & ($\arcsec$) & ($\arcsec$) & ($\degr$) & (mJy/beam) \\
\hline
Natural   & 1.54                         & 1.37                         & 47.9   & 0.17           \\
Natural+UVT (5\arcsec)  & 5.93                         & 4.71                         & -34.3  & 0.35           \\
Natural+UVT (5\arcsec)$^{*}$  & 22.0                       & 22.0                         & 0  & 1.20           \\
\textbf{Briggs ($r=0.5$)} & 1.26 & 1.16 & 44.7 & 0.18 \\
\hline
\end{tabular}
\tablefoot{UVT: uv-tapered images, using the tapering size in parenthesis.
\tablefoottext{*}{Convolved to reach a beam size of 22\arcsec}}
\end{table}

\subsubsection{CO(2-1) line\label{almalinecube}}

We perform the \texttt{clean} standard routine in the combined 12m and 7m data, using a mask based on the APEX CO(2$-$1) data described in \S \ref{apexlinecube}. We use briggs weighting with a robust parameter of 0.5. The reduced CO(2$-$1) cube has a FOV of 3'$\times$3', a velocity resolution of 0.5 km\,s$^{-1}$, an angular resolution of 1\farcs22$\times$1\farcs15 (PA$=$42\degr) and variable rms through out the line cube with values ranging from 22 and 36 mJy/beam per channel. 

This ALMA CO(2$-$1) cube is combined with the CO(2$-$1) APEX line cube as described in \S \ref{linecube}.

\subsection{APEX observations}

We use observations of Magellanic Bridge A performed with the APEX telescope\footnote{APEX is a collaboration between the Max-Planck-Institut fur Radioastronomie, the European Southern Observatory, and the Onsala Space Observatory.}, a 12m diameter telescope located in Llano de Chajnantor, Chile. These observations consist of an 870~$\mu$m continuum image, which we describe in \S \ref{laboca}, and a CO(2$-$1) line emission cube, described in \S \ref{apexlinecube}.

\subsubsection{0.87 mm continuum \label{laboca}}

We use 870 $\mu m$ continuum observations of Magellanic Bridge A obtained with the Large APEX Bolometer Camera (LABOCA) on the APEX telescope. LABOCA is an array of bolometers with a central frequency of 345 GHz, a bandwidth of 60 GHz 
and a total FOV of 11\farcm4 \citep{siringo2009}. 

We use the 870 $\mu m$ continuum image of Magellanic Bridge A presented in \citet{verdugo2012}. The Magellanic Bridge A data belongs to project C-086.F-0679A-2010 (PI M. Rubio). The observations were done in August and October, 2010, with a precipitable water vapour (pwv) content between 0.2 and 0.9 mm. The total integration time of the LABOCA observations was 3.94 hours. The reduction of this continuum image was done with the Bolometric Array Analysis Software (BoA)\footnote{BoA is a data reduction package developed specially to process data acquired with bolometer arrays at APEX telescope. The primary goal is to handle data observed with LABOCA}. The reduction required two calibrations: first, an opacity ($\tau$) calibration, using the $\tau$ values obtained from pwv measurements done with the APEX radiometer, and second a flux calibration based on Neptune and Uranus flux observations. The flux calibration has an estimated error of 20\%. The final continuum image was generated through an iterative reduction process: 

\begin{itemize}
    \item A first reduction was done with the standard BoA reduction script optimized for weak sources,
    \item Then, two iterations of the reduction code were done where pixels with S/N over 2.5 were flagged, in order to get rid of extra noise,
    \item Finally, the reduction code was done 6 times using a mask to detect emission over S/N of 2.5, subtract it, reduce the map without the source and add the source again before the next iteration begins.
\end{itemize}
 
The final image is presented in the bottom right box of Fig. \ref{fig:contprogression}: it has a beam size of 22\farcs4 (6.4 pc), a FOV of $11\farcm4\times11\farcm4$ and an rms of 5 mJy/beam, which is consistent with the integration time and weather conditions at the time of observation.

\subsubsection{CO(2-1) line \label{apexlinecube}}

We use CO(2$-$1) line observations, performed using the On-The-Fly mapping technique with the APEX$-$1 receiver of the Swedish Heterodyne Facility Instrument (SHeFI) at the APEX telescope. The receiver has a spectral range of 213 $-$ 275 GHz. These observations were made on June, 2014, under project C-093.F-9711A-2014 (PI M. Rubio), with pwv between 1.0 and 1.6 mm. The total integration time towards Magellanic Bridge A was 1.28 hours. The mapped area consists of a square of $5\farcs9\times5\farcs9$, covering the central region of the infrarred emission detected by the Herschel telescope (see Fig. \ref{fig:contprogression}). 

We reduce the APEX CO(2$-$1) line cube through the standard procedure of the CLASS software, Gildas\footnote{http://www.iram.fr/IRAMFR/GILDAS/}. The antenna temperature $T^*_{A}$ is delivered by APEX corrected for atmospheric attenuation\footnote{http://http://www.apex-telescope.org/observing/ (APECS user manual, Revision 4.1 June 5th, 2020)}. We scale $T^*_{A}$ to the main beam brightness temperature using $T_{mb} = T^{*}_{A}/\eta_{mb}$, with a main beam efficiency $\eta_{mb} = 0.72$ for APEX$-$1 \citep{Vassilev2008}. The resulting APEX CO(2$-$1) cube has a spatial resolution of 28\farcs7, a velocity channel spacing of 0.125 km\,s$^{-1}$ and an rms of 5.4 Jy/beam (150 mK) per channel. This line cube is combined with the resulting ALMA CO(2$-$1) line cube and described in \S \ref{linecube}.

\subsection{Combined ALMA \& APEX CO(2-1) observations\label{linecube}}

We combine the 12m and 7m ALMA CO(2$-$1) line cube described in \S \ref{almalinecube}, with the single dish APEX CO(2$-$1) line cube from \S \ref{apexlinecube}. To combine the observations, we use the \texttt{feather} task from the CASA software, which performs the combination in Fourier space. The final combined CO(2$-$1) datacube has a a FOV of 3\arcmin$\times$3\arcmin, a spatial resolution of $1\farcs22\times1\farcs15$ ($\approx0.3$ pc) and spectral resolution of 0.5 km\,s$^{-1}$. This CO(2$-$1) line cube has a variable rms, with values between 30 and 36 mJy beam$^{-1}$ per channel depending on position.

\subsection{Complementary data\label{sec:complementarydata}}

We use Herschel observations from the HERITAGE Herschel key project \citep{meixner2013}, consisting of maps of the SMC at 100, 160, 250, 350 and 500\,$\mu$m. We extract a square region of $11\arcmin\times11\arcmin$ from these maps, centered at $\alpha=$1:43:50, $\delta=$-74:32:23 (FK5, J2000), to work with the continuum images centered at Magellanic Bridge A. The angular resolution of the 100, 160, 250, 350 and 500 $\mu$m images is 9\arcsec, 14\arcsec, 22\arcsec, 30\arcsec and 43\arcsec, respectively, and the rms noise of each image is 9, 6, 0.6, 0.3 and 0.2 MJy/sr, respectively. The flux calibration of these images has an associated uncertainty of 10\% and 20\% for the 100 and 160 $\mu$m images, respectively, and $\sim$8\% for the 250, 350 and 500 $\mu$m images. We also use the 160\,$\mu m$ Spitzer image, obtained from the Multiband Imaging Photometer for Spitzer (MIPS) for the SAGE-SMC Spitzer Legacy Program \citep{gordon2011}, and extracted the region containing Magellanic Bridge A in the same way. The Spitzer image has a resolution of 40\arcsec, an rms of 4 MJy/sr and flux calibration uncertainty of $\sim$10\%. All these images were taken from the NASA/IPAC Infrarred Science Archive\footnote{https://irsa.ipac.caltech.edu/frontpage/}.

We use an H$\alpha$ image of the SMC from the Southern H$\alpha$ Sky Survey Atlas (SHASSA) \citep{gaustad2001}, which contains the Magellanic Bridge. The survey covers 13$\degr$ of the southern sky and has a resolution of about $0\farcm8$. We extract the region containing Magellanic Bridge A source the same way as with the Hershel images. 


\section{Millimeter and submillimeter emission from Magellanic Bridge A\label{sec:duststudies}}

In this section, we study the Magellanic Bridge A dust emission through the millimeter and submillimeter continuum images described in \S \ref{almacont}, \S \ref{laboca} and \S \ref{sec:complementarydata}. We first describe the sources present in the 870~$\mu$m continuum image. Then, we construct the spectral energy distribution (SED) of Magellanic Bridge A dust using the continuum images and subtracting all contributions that do not come from dust emission (\S \ref{SED}). We obtain the dust properties of Magellanic Bridge A by adjusting a Modified Blackbody (MBB) model to the SED (\S \ref{sec:model}). We compare the resulting MBB model with the dust emission derived from the LABOCA 870~$\mu$m and ALMA 1.3 mm continuum images (\S \ref{sec:excess}). Finally, we use the properties obtained from the model to calculate the total gas mass using dust emission (\S \ref{gasmasses}).

We present three different ALMA 1.3mm continuum images using different weights and tapered as explained in \S \ref{almacont} and no continuum source is detected in these ALMA continuum images, as seen in Fig. \ref{fig:almacont}. There is a hint of emission in the uv-tapered ALMA image convolved to 22\arcsec, but this emission is not coincident with Magellanic Bridge A in the continuum images from Herschel, Spitzer or LABOCA, so it might be related to antenna noise generating some artifacts.  We use the natural weighted, uv-tapered image to obtain an upper limit for the flux density at 1.3~mm (see \S \ref{SED}).

\begin{figure*}[ht]
\centering
\includegraphics[width=\hsize]{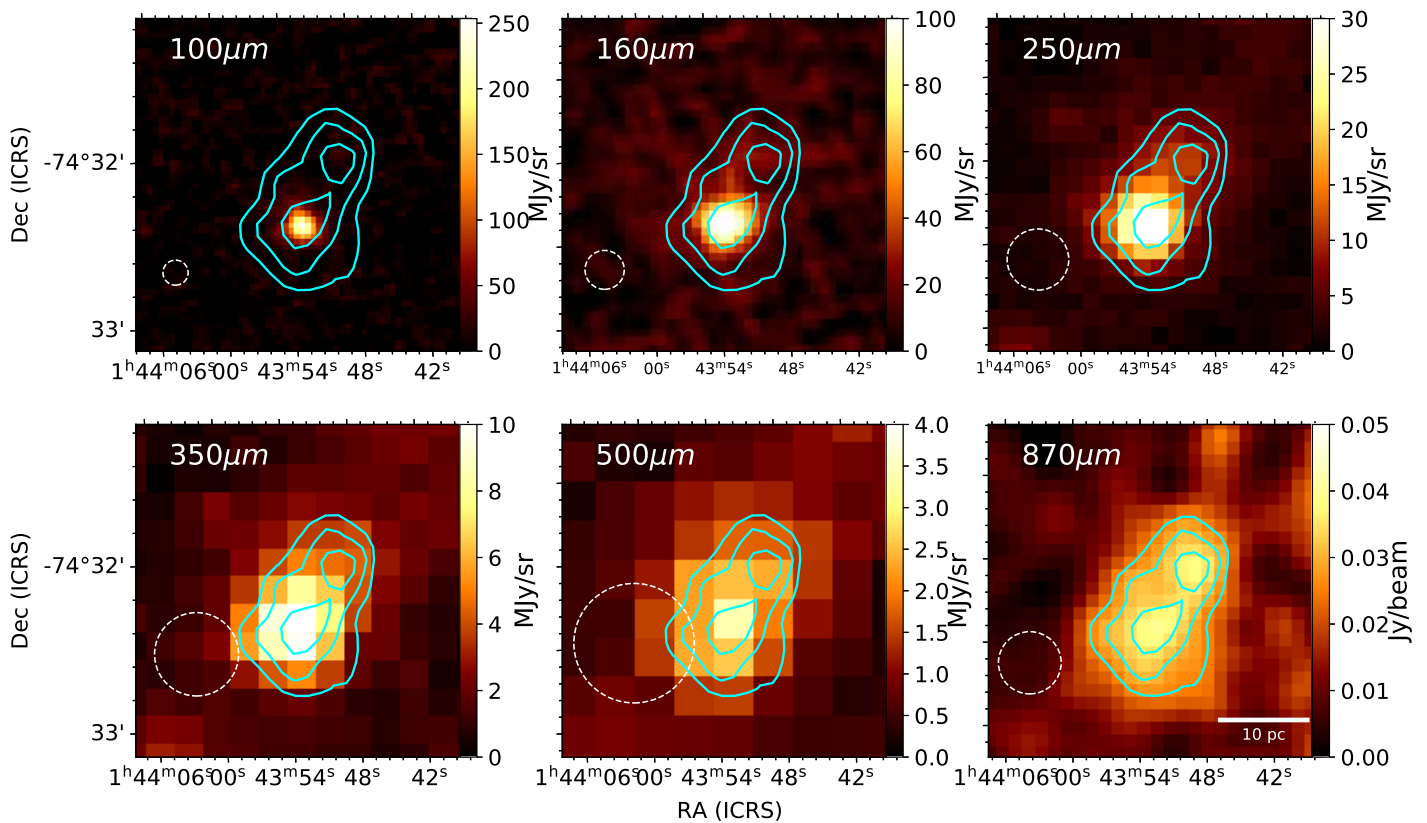}
\caption{Herschel continuum images at 100, 160, 250, 350 and 500~$\mu$m, together with the LABOCA continuum image at 870~$\mu$m, of Magellanic Bridge A. The cyan contours correspond to the LABOCA 870~$\mu$m continuum image at 25 (5$\sigma$), 30 (6$\sigma$) and 35 mJy/beam (7$\sigma$). The dashed circles indicate the beam sizes for each image, which correspond to a FWHM of 9\arcsec (100 $\mu$m), 14\arcsec (160 $\mu$m), 22\arcsec (250 $\mu$m), 30\arcsec (350 $\mu$m), 43\arcsec (500 $\mu$m) and 22\farcs4 (870~$\mu$m). The scale bar on the bottom left of the lower right panel indicates a 10 pc length. \label{fig:contprogression}}
\end{figure*}

We detect continuum emission from Magellanic Bridge A at 870~$\mu$m. The APEX continuum image is shown in the bottom right panel of Fig. \ref{fig:contprogression}. The source is also detected in the Herschel 100, 160, 250, 350 and 500~$\mu$m images. We can resolve two peaks with similar intensities ($\sim 40$ mJy/beam) and separated by $\sim27\arcsec$ (8 pc), which is near the limit of the APEX resolution. We identify these two sources as North and South throughout this paper. In the Herschel 100 and 160~$\mu$m images, the northern peak is not seen while the southern source can be seen in all images. Unlike the APEX 870~$\mu$m continuum, the 350 and 500 $\mu$m lower resolution images do not separate well the northern component, but it can be seen as an elongated structure. For comparison, we plot the APEX contours over the Herschel continuum images in Fig. \ref{fig:contprogression}. In the following section, we characterize the dust emission towards Magellanic Bridge A and we also characterize the emission coming from sources North and South found in the APEX 870~$\mu$m image.
 
\subsection{Spectral energy distribution of Magellanic Bridge A\label{SED}}

We build the SED for Magellanic Bridge A using Herschel, Spitzer, APEX and ALMA continuum images. We also build the SED for the North and South sources using Herschel emissions only at 100, 160 and 250 $\mu$m, together with the LABOCA and ALMA continuum images. To construct each SED, we perform aperture photometry at each wavelength where emission was detected to obtain the fluxes. Then, we subtract the contributions that are not thermal emission. We determine an upper flux limit in the ALMA 1.3 mm continuum image, as no emission is detected. 

In the following, we describe how we construct the SED and how we determine the contribution from other emission mechanisms.

\subsubsection{Flux measurements\label{aphot}}

We obtain the fluxes in the Herschel, Spitzer and LABOCA continuum images of Magellanic Bridge A using aperture photometry. We first convolve all images to a common resolution of 43\arcsec, which corresponds to the resolution of the Herschel 500~$\mu$m image, as it has the poorest resolution of all our images. Convolving results in a change in the rms of each image: the new rms are 2.0, 5.0, 0.5 and 0.3 MJy/sr for the Herschel 100, 160, 250 and 350 $\mu$m images, respectively. The convolved LABOCA image has an rms of 26 mJy/beam. We measure the total emission coming from Magellanic Bridge A in an aperture with a radius $r=50\arcsec$, centered at $\alpha=$ 1:43:52.68, $\delta=$ -74:32:20.23 (FK5, J2000). For all images, we subtract the sky using an annulus centered at the same position as the aperture, with inner radius $r_{in}=2\farcm5$ and outer radius $r_{out}=3\farcm0$ for Magellanic Bridge A. We do not apply an aperture correction to the photometry results of Magellanic Bridge A because the aperture encloses the source completely and the aperture area is almost 4 times the beam size. The total flux error is the sum in quadrature of the flux calibration error for the measured flux and the photometric error $\epsilon$:
\begin{equation}
    \epsilon=\sqrt{N} \sigma\label{eq:photerror}
\end{equation}
where N is the number of beams inside the aperture and $\sigma$ is the rms of the image in Jy/beam. The fluxes measured for Magellanic Bridge A are given in the first row of Table \ref{fluxresults}. The 870~$\mu$m and 1.3 mm emission have been corrected for free-free emission and CO (2$-$1) line contribution at the respective wavelength. These corrections are further explained in \S \ref{sec:freefreeandco}. In the Spitzer 160 $\mu$m image (not included in Table \ref{fluxresults}), we measure a flux density $S_{160\mu m}=1009\pm101\pm386$, where the first error corresponds to the flux calibration error and the second is the photometric error. 

In the case of Magellanic Bridge A North and South, we only use the 100, 160 and 250~$\mu$m images from Herschel and the LABOCA 870~$\mu$m continuum. We do not use the 350 and 500~$\mu$m images because the North and South sources cannot be separated due to their resolution (their beam FWHM are larger than 22\arcsec.). We convolve the Herschel images to a common resolution of 22\arcsec to measure their fluxes. The rms after convolution are 2.2 and 5.5 MJy/sr for the Herschel 100 and 160~$\mu$m images, respectively. For the photometry, we use a circular aperture with a common radius $r=11\arcsec$, centered at $\alpha=$ 1:43:49.0, $\delta=$ -74:32:00.2 (FK5, J2000) for source North and at $\alpha=$ 1:43:53.8, $\delta=$ -74:32:24.5 (FK5, J2000) for source South, which correspond to the peak positions in the LABOCA image. In this case, since the sources are not resolved and the photometry is performed within the FWHM of the spatial beam, we apply an aperture correction factor of 2 in the flux measurements. The total flux error is also calculated as the sum in quadrature of the associated flux calibration error and photometric error.

Since there is no emission detected in the ALMA 1.3 mm image, we determine an upper limit for the flux at 1.3 mm for sources Magellanic Bridge A, North and South, based on the ALMA continuum image sensitivity. To do this, we use a value 3 times the photometric error $\epsilon$. For Magellanic Bridge A, we obtain a photometric error $\epsilon =5.5$ mJy inside an $r=50\arcsec$ circular aperture, which gives a 1.3~mm flux upper limit for Magellanic Bridge A of 16.5 mJy. The photometric error for the North and South sources is $\epsilon = 2.4$ mJy in a $r=11\arcsec$ circular aperture, after applying an aperture correction of a factor of 2. This gives a 1.3~mm flux upper limit of 7.2 mJy for sources North and South.

The fluxes measured for each source and wavelength, together with the flux upper limit at 1.3 mm, are summarized in Table \ref{fluxresults}. The 1.3 mm and 870~$\mu$m fluxes listed correspond to dust emission only at these wavelengths as free-free emission and CO line contributions have been estimated and subtracted (see \S \ref{sec:freefreeandco}). Using these fluxes, we build the SED shown in Fig. \ref{fig:SED}.

\begin{table*}[ht]
\centering
\caption{Dust emission flux densities in Magellanic Bridge A, North and South} 
\label{fluxresults}
\begin{tabular}{lllllllll}
\hline\hline 
Source  & $R_{ap}$ & $S_{100\mu m}$     & $S_{160\mu m}$       & $S_{250\mu m}$    & $S_{350\mu m}$    & $S_{500\mu m}$    & $S_{870\mu m}^{\dagger}$    & $S_{1.3mm}^{\ddagger}$ \\
        & (\arcsec) & (mJy)              & (mJy)                & (mJy)             & (mJy)             & (mJy)             & (mJy)             & (mJy)       \\\hline
MagB. A$^{*}$ & $50 $      & 952$\pm95\pm151$ & 1565$\pm313\pm511$ & 856$\pm68\pm48$ & 422$\pm34\pm33$ & 187$\pm15\pm23$ & 257$\pm52\pm28$ & $<16 $      \\
North$^{**}$  & $11$       & 131$\pm13\pm49$  & 305$\pm61\pm120$   & 139$\pm11\pm12$ & …                 & …                 & 46$\pm9\pm5$    & $<7$        \\
South$^{**}$   & $11$       & 678$\pm68\pm48$  & 838$\pm168\pm120$  & 449$\pm36\pm13$ & …                 & …                 & 56$\pm11\pm5$   & $<7$  \\\hline     
\end{tabular}
\tablefoot{We express the errors of our measurements with the flux calibration error first (Herschel errors obtained from \citealt{meixner2013}) and the photometric error last.
\tablefoottext{$\dagger$}{Flux at 870~$\mu$m has the free-free emission and CO(3$-$2) line contributions subtracted.}
\tablefoottext{$\ddagger$}{Flux at 1.3 mm has the free-free emission contribution subtracted.}
\tablefoottext{*}{In Magellanic Bridge A, all images are convolved to a common resolution of 43\arcsec.}
\tablefoottext{**}{In sources North and South, all images are convolved to a common resolution of 22\arcsec.}
 }
\end{table*}

 \subsubsection{Free-free and CO(3-2) contribution \label{sec:freefreeandco}}

The flux measurement from the APEX 870~$\mu$m image contains emission of three different physical processes: 1) thermal dust continuum emission, 2) free-free (bremsstrahlung) emission from ionized gas, and 3) a contribution from molecular lines in the passband (\element[][12]{CO}(3$-$2) in particular, which is the brightest). At 1.3 mm, there is also free-free emission and a molecular line (\element[][12]{CO}(2$-$1)). We are interested in the dust continuum emission, therefore we need to calculate and remove the other two contributions to the measured flux. The CO(2$-$1) line emission is already removed when the ALMA 1.3 mm continuum image is constructed (see \S \ref{almacont}), so we determine the contribution of free-free to both 870~$\mu$m and 1.3 mm continuum emission and the contribution of \element[][12]{CO}(3$-$2) line emission in the 870~$\mu$m continuum image.

We determine the free-free contribution to continuum emission using H$\alpha$ emission, as the intensity of the H$\alpha$ line traces the warm ionized gas component of the ISM and, therefore, is proportional to the free-free emission \citep[see the Appendix of][]{hunt2004}. To convert H$\alpha$ flux density into free-free emission $S_{\nu}^{ff}$ in mJy, we use the following formula:

\begin{equation}
S_{\nu}^{ff} = 1.16 \left( 1+\frac{n(\mathrm{He}^+)}{n(\mathrm{H}^+)}\right) \left(\frac{T_\mathrm{e}}{10^4}\right)^{0.62} \nu^{-0.1} \left( \frac{S_{\alpha}}{10^{-12}}\right)\label{eq:ff}
\end{equation}
Where $S_{\alpha}$ is the intensity of H$\alpha$ in erg\,cm$^{-2}$\,s$^{-1}$, $T_\mathrm{e}$ is the gas electron temperature in K, $n(\mathrm{He}^+)$ and $n(\mathrm{H}^+)$ are the number density of Helium and Hydrogen ions, respectively, and $\nu$ is the frequency where we need to determine the free-free emission in GHz. To obtain the intensity of H$\alpha$ emission in Magellanic Bridge A, we use the H$\alpha$ image from SHASSA \citep{gaustad2001}, which is in deciRayleighs (dR). We first transform the map units to Rayleighs (R) dividing by 10 and then to erg\,cm$^{-2}$\,s$^{-1}$\,sr$^{-1}$ using $1\mathrm{R}=h\nu\times 10^6/(4\pi)=2.41\times 10^{-7}$ erg cm$^{-2}$ s$^{-1}$ sr$^{-1}$, where $\nu$ is the frequency corresponding to the H$\alpha$ line ($4.57\times10^{14}$ Hz). The H$\alpha$ measurements obtained through aperture photometry in the same positions as in section \ref{aphot} are $(1.2\pm0.2)\times10^{-13}$ erg cm$^{-2}$ s$^{-1}$ for for Magellanic Bridge A, and $(2.3\pm0.3)\times10^{-14}$ erg cm$^{-2}$ s$^{-1}$ and $(2.4\pm0.4)\times10^{-14}$ erg cm$^{-2}$ s$^{-1}$ for the North and South sources, respectively. These values consider an extinction $A_V = 0.140$, obtained for the position of Magellanic Bridge A \citep{schlafly2011}. We use $T_\mathrm{e}\sim1.7\times10^4$ K, the upper limit for the gas temperature reported in \citet{lehner2001}, and $n(He^+)/n(H^+)\sim 0.08$, which is the value estimated for low metallicity sources like the Magellanic Clouds \citep{hunt2004}. 

After applying Equation \ref{eq:ff} using $\nu=344.8$ GHz (870~$\mu$m), we obtain a free-free contribution of 117$\pm$22 $\mu$Jy for Magellanic Bridge A, 22$\pm$4 $\mu$Jy for cloud North and 23$\pm$4 $\mu$Jy for cloud South. These values represent $\lesssim0.05\%$ of the total continuum emission. When we use Equation \ref{eq:ff} with $\nu=230.8$ GHz (1.3 mm), we obtain a free-free contribution of 122$\pm$23 $\mu$Jy in Magellanic Bridge A, 23$\pm$4 $\mu$Jy in cloud North and 24$\pm$4 $\mu$Jy in cloud South. The values for each source and wavelength are listed in Table \ref{freefreeresults}. The fluxes presented in Table \ref{fluxresults} have these contributions subtracted.

\begin{table}[ht]
\centering
\caption{Free-free emission for each source at 870~$\mu$m and 1.3 mm, obtained through aperture photometry}  
\label{freefreeresults}
\begin{tabular}{lll}
\hline\hline
Source     & $S_{0.87mm}^{ff}$ (mJy) & $S_{1.3mm}^{ff}$ (mJy) \\
\hline
Magellanic Bridge A$^*$  &0.117$\pm$0.022         & 0.122$\pm$0.023           \\
North$^{**}$      & 0.022$\pm$0.004              & 0.023$\pm$0.004              \\
South$^{**}$      & 0.023$\pm$0.004               & 0.024$\pm$0.004          \\
\hline
\end{tabular}
\tablefoot{
\tablefoottext{*}{Aperture photometry done in a circle with a radius of 50\arcsec.}
\tablefoottext{**}{Aperture photometry done in a circle with a radius of 11\arcsec}}
\end{table}

We obtain the \element[][12]{CO}(3$-$2) molecular line contribution to the LABOCA continuum flux $S_{CO(3-2)}$ in Magellanic Bridge A using Equation 5 of  \citet{drabeck2012}. We calculate the total flux density present in Magellanic Bridge A from the peak integrated intensity $I_{CO(3-2)}$ using:
\begin{equation}
     S_{^{12}CO(3-2)} = \frac{2k \nu^3}{c^3 \Delta \nu_{bol}} \Omega I_{^{12}CO(3-2)}
\end{equation}
where $I_{CO(3-2)}$ is in K km s$^{-1}$, $\Delta \nu_{bol}$ is the bandpass width in Hz, $\nu$ is the frequency of the \element[][12]{CO}(3$-$2) line in Hz, $\Omega$ is the source area in sr and $S_{CO(3-2)}$ is in Jy. We approximate the LABOCA spectral response to a function that is constant over the LABOCA bandpass, so that $g(\nu)_{line}/\int g(\nu)d\nu=1/\Delta \nu_{bol}$. We use 
$I_{^{12}CO(3-2)}=0.93\pm 0.07$ K km s$^{-1}$, the value for the peak $I_{^{12}CO(3-2)}$ in Magellanic Bridge A given in Table 2 of \citet{muller2014}, and its area $\Omega$ corresponds to one single ASTE pointing, with $FWHM=22\arcsec$. The resultant \element[][12]{CO}(3$-$2) line contribution to the continuum emission at 870~$\mu$m is $S_{CO(3-2)}=0.85\pm 0.06$ mJy. This value is subtracted to the total flux in Magellanic Bridge A at 870 microns, but not to the North and South sources, as there is only one source detected in \element[][12]{CO}(3$-$2). Nevertheless, the estimated \element[][12]{CO}(3$-$2) contribution is less than 2\% of the 870~$\mu$m flux in sources North and South.

We use the APEX 870~$\mu$m flux measurement with the free-free and \element[][12]{CO}(3$-$2) line emission contribution subtracted, the ALMA 1.3 mm flux upper limit with free-free contribution subtracted, and the FIR flux densities obtained from Herschel and Spitzer images to construct the SED of Magellanic Bridge A and the North and South sources. The three SEDs are plotted in Fig. \ref{fig:SED}.


\subsection{SED modeling\label{sec:model}}

We use the SED constructed in \ref{sec:freefreeandco} for Magellanic Bridge A to obtain the dust temperature and dust emissivity index of the region. We also use the SEDs of sources North and South to obtain their dust temperatures. Each SED is modeled assuming that the emission comes from a MBB, which consists of the Planck function multiplied by an emissivity that depends on frequency and dust properties. We derive the physical properties of these dust sources using the parameters of the MBB that best fit each source SED.

At millimeter and submillimeter wavelengths, we can assume that dust emission is optically thin, which means we can write the power per unit area, frequency and solid angle as $I_{\nu} = \tau_{\nu} B_{\nu}(T_{d})$, where $\tau_{\nu}$ is the dust optical depth and $B_{\nu}$ is the Planck's law value at the dust temperature $T_d$. $\tau_{\nu}$ can be defined through the dust's absorption coefficient (in cm$^2$ g$^{-1}$) as $\tau_{\nu}=\kappa_d(\nu)\Sigma_{d}$, where $\Sigma_{d}$ is the surface density of the source in g cm$^{-2}$. Replacing $\tau_{\nu}$ and integrating $I_{\nu}$ by the solid angle we get:

\begin{equation}
S_{\nu} = \Omega \kappa_d(\nu) \Sigma_d B_{\nu}(T_d)
\end{equation}
where $\Omega$ is the solid angle of the source. 

In the MBB model, we write the absorption coefficient $\kappa_d(\nu)$ as a function of the spectral emissivity index $\beta$ as $\kappa_d(\nu) = \kappa_d(\nu_0)(\nu/\nu_0)^{\beta}$, where $\kappa_d(\nu_0)$ is the absorption coefficient at the reference frequency $\nu_0$. Replacing $\kappa_{d}(\nu)$ and reorganizing the constant values $\Omega$, $\Sigma_d$, $\kappa_d(\nu_0)$ and $\nu_0$ as $C$, the flux density is a function of frequency, $\beta$, the dust temperature and the constant C \citep{hildebrand1983}: 

\begin{equation}
S_{\nu} = C \nu^{\beta} B_{\nu}(T_d)\label{eq:mbbmodel}
\end{equation}

\begin{figure*}[ht]
   \centering
   \includegraphics[width=\hsize]{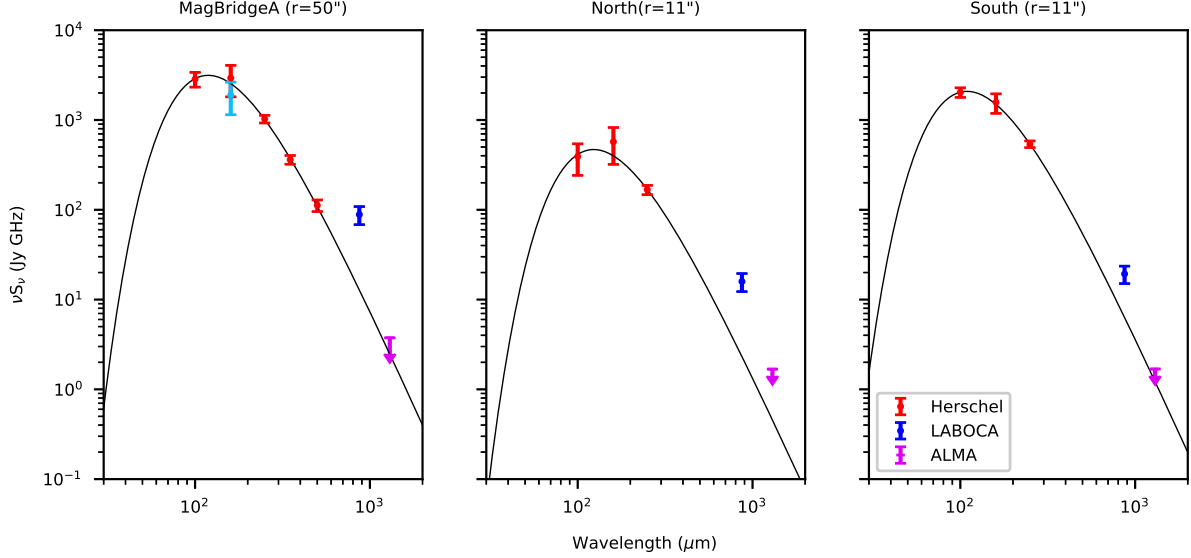}
   \caption{SED for Magellanic Bridge A and the North and South sources. The points correspond to the energy $\nu S_{\nu}$ derived from the measured flux density $S_{\lambda}$, obtained using aperture photometry. Red points correspond to $\nu S_{\nu}$ obtained using Herschel data. Dark blue points correspond to the energy $\nu S_{\nu}$ obtained using the LABOCA image. The purple arrow corresponds to the $\nu S_{\nu}$ upper limit at 1.3 mm using the ALMA continuum image. The left panel shows the Magellanic Bride A SED, for which we used the 100, 160, 250, 350 and 500 $\mu$m Herschel continuum images, the 160 $\mu$m Spitzer continuum image, the 870~$\mu$m (LABOCA) and 1.3 mm (ALMA) continuum images with a common resolution of 43\arcsec. We measure the flux from Magellanic Bridge A inside a circular aperture of radius $r=50\arcsec$. The North and South sources SEDs are shown in the last two panels, constructed with the Hershel 100, 160 and 250 $\mu$m continuum images, the 870~$\mu$m (LABOCA) and 1.3 mm (ALMA) continuum images at a common resolution of 22\arcsec. We measure the fluxes from sources North and South in a circular aperture of radius $r=11\arcsec$. The MBB model that best fits the Herschel points for each source is shown as a black curve. The parameters of each curve are in Table \ref{tab:parameters}. \label{fig:SED}}
    \end{figure*}

We fit the MBB model described by Equation \ref{eq:mbbmodel} for Magellanic Bridge A using the 100, 160, 250, 350 and 500 $\mu$m fluxes, obtained from Herschel images in a $r=50\arcsec$ radius aperture (see \S \ref{aphot}). We assume a single dust component and single $\beta$ dust emissivity index. We have 3 free parameters in the model: $C$, $\beta$ and $T_d$. We use $\chi^2$ minimization to determine $C$, $\beta$ and $T_d$ that best fit the SED, using the scipy.optimize package \citep{scipy2020} function curve\_fit. The MBB that best fit the SED of Magellanic Bridge A gives a dust temperature $T_d=22.4\pm3.4$ K and an emissivity index $\beta=1.4\pm0.5$, listed in Table \ref{tab:parameters}. In Fig. \ref{fig:SED}, the solid black line in the left panel represents the best fit MBB for the SED of Magellanic Bridge A. 

We fit the MBB model for sources North and South using the 100, 160 and 250 $\mu$m fluxes only, because the 350 and 500 $\mu$m beams do not resolve two sources and have poorer resolution than 22\arcsec (the 870~$\mu$m beam). For these sources, we assume that the dust emissivity is the same as the one found for Magellanic Bridge A, and therefore fix $\beta=1.4$ for both sources. We note that assuming a $\beta$ value for these sources influences the dust temperature obtained, as there is an inverse correlation between $\beta$ and $T_d$ \citep{shetty2009}. We find the parameters $C$ and $T_d$ that best fit the SEDs with $\chi^2$ minimization. The temperature obtained for source South is $T_d=24.5.4\pm2.2$ K and source North $T_d=21.7.4\pm1.1$ K, parameters which are listed in Table \ref{tab:parameters} and the best fit curves are plotted with a black line in In Fig. \ref{fig:SED}. The best fit results indicate that the North source is colder than the South source by $\sim3$ K.

\begin{table*}[ht]
\centering
\caption{Gas masses and best fit MBB parameters to the clouds SED}
\begin{tabular}{llllll}
\hline\hline
Source     & C  (sr Hz$^{-\beta}$)   & $\beta$      & $T_d$ (K)          & $\chi ^2$ & $M_{gas}$ (10$^3$ M$_{\sun}$)$^{\ddagger}$ \\
\hline
MagBridgeA$^*$ & $(1.9\pm23.4)\times10^{-29}$  & $1.4 \pm0.5$ & $22.4 \pm 3.4$ & 0.27      & $6.8\pm1.5$                 \\
North$^{**}$       & $(9.0\pm3.1)\times 10^{-30}$ & $1.4^\dagger$  & $21.7\pm2.2$   & 0.5      & $1.3\pm0.5$            \\
South$^{**}$       & $(2.1\pm0.4)\times10^{-29}$ & $1.4^\dagger$  & $24.5\pm1.1$  & 0.04     & $2.9\pm 0.5$                \\
\hline
\end{tabular}
\tablefoot{\tablefoottext{*}{Fluxes measured using aperture photometry radius of 50\arcsec.}
\tablefoottext{**}{Fluxes measured using aperture photometry radius of 11\arcsec.}
\tablefoottext{$\dagger$}{$\beta$ values obtained for Magellanic Bridge A MBB fit.}
\label{tab:parameters} \tablefoottext{$\ddagger$}{Gas masses derived from the MBB models that best fit the FIR SEDs of each source.}}
\end{table*}

\subsection{Submillimeter excess\label{sec:excess}}

We define the submillimeter and millimeter excesses $E(\lambda)$ as the ratio between the measured flux and the predicted flux from the SED model at $\lambda=870$ $\mu$m and 1.3 mm. We calculate these excesses for each source and summarize our results in Table \ref{tab:excesses}.

The submillimeter excess found at 870~$\mu$m for Magellanic Bridge A is $E(870\,\mu m)=6.9\pm1.6$ inside an circular aperture with $r=50\arcsec$. When we separate Magellanic Bridge A into the North and South source, this submillimeter excess is also present. For the North source, $E(870\,\mu m)=6.7\pm1.6$, while in the South is $E(870\,\mu m)=3.0\pm1.4$, both considering a photometry aperture radius $r=11\arcsec$. Excess emission at 870~$\mu$m has been detected using LABOCA observations in other low metallicity galaxies \citep{galametz2009,galametz2014,hermelo2016}, as well as with SCUBA \citep{galliano2003, galliano2005,bendo2006}.

Since we do not detect our sources in the 1.3 mm ALMA continuum, we calculate how much higher are the flux upper limits calculated in \S \ref{aphot} than the predicted flux densities at 1.3 mm. We derive an expected flux using the MBB model that best fits the FIR SED (see Table \ref{tab:parameters}). We obtain an expected flux density of 10.7 mJy for Magellanic Bridge A in an aperture radius $r=50\arcsec$, and 2.0 mJy for the North source and 5.3 mJy for the South source within an aperture of radius $r=11\arcsec$.
The flux upper limit at 1.3 mm is 16.5 mJy, <1.5 times the expected flux density using the MBB model fitted to the FIR fluxes for Magellanic Bridge A. For the individual North and South sources, the upper limits from the ALMA continuum image at 1.3 mm are 7.2 mJy, $<3.6$ times the expected flux density for source North, and 2.2 mJy, $<1.4$ for source South. These values are reported as $E(1.3\,mm)$ in Table \ref{tab:excesses}. In summary, our ALMA 1.3 mm data are on the sensitivity limit compatible with the MBB fit to the far-infrared observations, and show no excess at this wavelength. If the excess at 1.3 mm were similar to that at 870~$\mu$m, we would have clearly detected the source.

The fact that we do not observe emission at 1.3 mm when there is notable excess emission at 870~$\mu$m in our sources is puzzling. We have verified that the calibration of the LABOCA data is correct by performing independent reductions of the data by members of our team and the instrument support team, which result in very similar images. Also, we note that the morphology of the source at 870~$\mu$m in Fig. \ref{fig:contprogression} is consistent with emission at 250 to 500 $\mu$m continuum images, and the noise of the map is consistent with that expected for the integration time. In fact, we would have only detected the South source at $\sim3\sigma$ in the complex if there were no excess. Therefore, we have every expectation that the 870~$\mu$m excess is real. If this is so, there are two possibilities to explain the lack of a 1.3~mm detection: either the anomalous emission at 870~$\mu$m is only present at that wavelength, or dust emission is too extended and resolved out in the ALMA 1.3 mm observation. We note that it is very difficult to explain the 870~$\mu$m as resulting from an additional very cold dust component, as the lack of an excess at 500~$\mu$m would require it to have a dust temperature $T_d\sim3$ K. Through the inclusion of the Morita Array our ALMA observations have a maximum recoverable scale (MRS) of 30.9\arcsec, although our sensitivity substantially degrades close to that limit. We would expect to see continuum sources similar in size to the CO emitting clouds in this region (Fig. \ref{fig:CO2-1}) if their 1.3~mm flux had an excess over the MBB fit similar to the one we measure at 870~$\mu$m. If the sources are larger than the 22\arcsec used for the photometry, however, they would be at least partially resolved out and undetectable. Clearly future ALMA observations determining the sizes of the sources at 870~$\mu$m or deeper observations at 1.3 mm would be extremely useful to establish the nature of the excess.

\begin{table}[ht]
\centering
\caption{Submillimeter and millimeter excesses for Magellanic Bridge A and sources North and South}
\begin{tabular}{lll}
\hline\hline  
Source & E(870)  & E(1.3mm)       \\
\hline  
MagBridgeA &6.9$\pm$1.6 & <1.5 \\
North & 6.7$\pm$1.6 & <3.7 \\
South & 3.0$\pm$0.7 & <1.4 \\
\hline  
\end{tabular}\label{tab:excesses}
\tablefoot{Excess is defined as the ratio between the observed flux density and the flux density predicted by the best fit MBB model to the FIR dust fluxes in the SED.}
\end{table}

\subsection{Gas masses obtained from dust emission \label{gasmasses}}

We calculate the total gas masses associated to the 870~$\mu$m dust emission for Magellanic Bridge A predicted by the MBB model that best fits the FIR fluxes. We use the same method to obtain the gas mass from dust flux used in \citet{bot2010} and \citet{rubio2004}. The dust flux $S_{\nu}$ is related to the gas mass $M_{gas}$ through:

\begin{equation}
    M_{gas} = \frac{S_{\nu}D^2 \mu m_H}{\epsilon_d(\nu)B_{\nu}(T_d)}\label{eq:gasmass}
\end{equation}
where $S_{\nu}$ is the predicted flux at frequency $\nu$ by the MBB model that best fits the FIR data in Jy, $D$ is the distance to the source in cm (we use the estimated distance to the SMC, 60 kpc, \citealt{harries2003}), $\mu m_H$ is the gas weight per Hydrogen atom in gr, including the contribution of Helium\footnote{$\mu=1.36$}, $\epsilon_d(\nu)$ is the emissivity of dust per Hydrogen atom at frequency $\nu$ in cm$^2$ and $B_{\nu}(T_d)$ is the spectral radiance at a dust temperature $T_d$ in Jy sr$^{-1}$.

We use the $\epsilon_d(345GHz)$ value presented in \citet{bot2010}, $\epsilon_d(345GHz)=(3.94\pm0.05)\times10^{-27}$ cm$^2$, which assumes an absorption coefficient $\kappa_{345GHz}=1.26\pm0.02$ cm$^2$\,g$^{-1}$ and a dust-to-gas ratio 1/6 of the ratio in the solar neighborhood. This dust-to-gas ratio is similar to the ratios derived in Magellanic Bridge B and C \citep{gordon2009}.

For Magellanic Bridge A, North and South, we use the dust temperatures $T_d=22.4\pm3.4$ K, $T_d=21.7\pm2.2$K and $T_d=24.5\pm1.1$ K, respectively, obtained from the best fit MBB to the Magellanic Bridge A SED with $\beta=1.4$, as derived in \S \ref{sec:model}. 

The total gas mass are calculated using the 870~$\mu$m flux predicted from the MBB fit. For Magellanic Bridge A, the gas mass is $(6.8\pm1.5)\times 10^{3}$ M$_{\sun}$ and for the two separate clouds, North and South, their gas masses are $(1.3\pm0.5)\times 10^{3}$ M$_{\sun}$ and $(2.9\pm0.5)\times 10^{3}$  M$_{\sun}$, respectively. The difference between the sum of the gas masses of both sources and the 
gas mass of Magellanic Bridge A is because dust measurements for North and South sources are obtained in an aperture of radius $r=11\arcsec$, which only includes the peak emission of each cloud, while Magellanic Bridge A is measured inside an aperture of radius $r=50\arcsec$, which captures the complete emission from the source. We also estimated the gas masses using an absorption coefficient determined for dust emission at 160 $\mu$m by \citet{gordon2014}, $\kappa_{160\mu m}=9.6\pm2.5$ cm$^2$ g$^{-1}$ as the gas masses are dependant on the chosen $\epsilon_d(\nu)$, which depends on the calibration of $\kappa_{\nu}$. Using $\kappa_{160\mu m}$ and the same gas-to-dust ratio  $1/6$, we obtain a 
gas mass for Magellanic Bridge of $(9.6\pm4.4)\times10^3$ M$_{\sun}$, while for source North is $(1.7\pm0.8)\times10^3$ M$_{\sun}$ and for source South is $(3.9\pm1.4)\times10^3$ M$_{\sun}$. These values are within the uncertainties of those obtained using the \citet{bot2010} value. For the rest of this work, we use the total gas masses derived from the \citeauthor{bot2010} calibration, listed in the last column of Table \ref{tab:parameters}.

\section{Resolved CO(2-1) molecular clouds\label{molclouds}}

\begin{figure}[ht]
\centering
\includegraphics[width=\hsize]{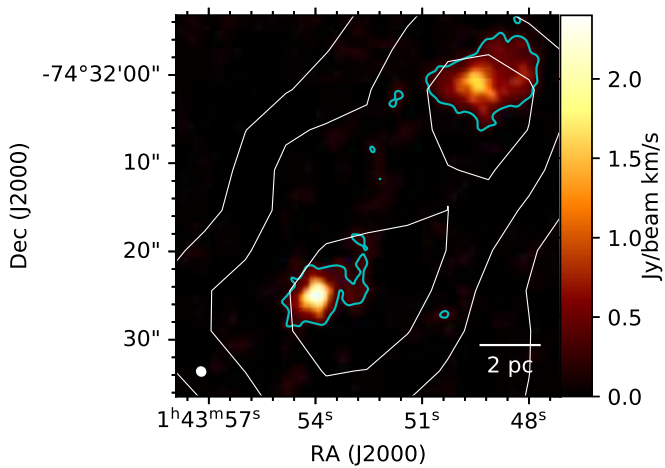}
\caption{\label{fig:CO2-1}Velocity integrated intensity image of the combined ALMA and APEX CO(2$-$1) cube of Magellanic Bridge A, integrated between 172 and 176.0 km s$^{-1}$. The cyan contours correspond to 5 times the $\sigma$ of the integrated image ( $\sigma=48$ mJy/beam km s$^{-1}$). The white contours correspond to the LABOCA 870~$\mu$m continuum image at 25, 30 and 35 mJy/beam, as in Figure \ref{fig:contprogression}. The white ellipse in the lower left corner represents the beam FWHM ($1.22\arcsec\times1.15\arcsec$). The scale bar on the lower right corner indicates a 2 pc length.}
\end{figure}

In this section, we study the CO(2$-$1) molecular line emission in Magellanic Bridge A. We produce the velocity integrated intensity image between 172 and 176 km\,s$^{-1}$ of the ALMA and APEX CO(2-1) line cube (see Fig. \ref{fig:CO2-1}). We measure an rms of 48 mJy\,beam$^{-1}$\,km\,s$^{-1}$ in this image, a value which is consistent with the calculated values of 42 and 50 mJy/beam km/s using the rms range of the velocity interval (see \S \ref{linecube}). We resolve two parsec-sized CO(2$-$1) clouds, which are located towards the North and South 870~$\mu$m continuum sources. We extract the spectra of each of these molecular clouds in \S \ref{sec:co21spectra} and determine the physical properties of the molecular clouds in this region in \S \ref{cloudprops}.

\subsection{CO(2-1) molecular clouds spectra\label{sec:co21spectra}}

We identify the molecular clouds in the ALMA and APEX CO(2$-$1) line cube and extract their spectra. We first identify potential sources in the velocity integrated intensity image as all emission greater than 3$\sigma$, being $\sigma$ the rms of the integrated image. We define the area of each source as the number of pixels inside the 3$\sigma$ contour. Then, we measure the second moments of emission along the major and minor axes of each source in the CO(2$-$1) integrated emission image. We only keep those sources whose minor second moment is larger than the beam size. The only sources that meet this criteria (survive this decimation) are the two brightest CO(2$-$1) sources seen in Figure \ref{fig:CO2-1}, which we call CO clouds from now on. 
These CO clouds spatially coincide with the dust sources characterized in \S \ref{sec:duststudies}, as seen in Figure \ref{fig:CO2-1}. The peak $I_{CO(2-1)}$ of these clumps in the CO(2$-$1) integrated image are 1.67$\pm$0.05 Jy beam$^{-1}$ km s$^{-1}$ in the North CO cloud and 2.81$\pm$0.05 Jy beam$^{-1}$ km s$^{-1}$ in the South CO cloud. 

\begin{table*}[ht]
\centering
\caption{\label{tab:COspectrachar}Characteristics of the clouds identified in the CO(2$-$1) line cube.}
\begin{tabular}{llllllll}
\hline\hline
CO Source & RA  & DEC & $A$ & Peak $I_{CO(2-1)}^{(a)}$  & $v_{LSR}$ $^{(b)}$  & $\Delta v$ (FWHM) $^{(b)}$  & $S_{CO(2-1)}\Delta v$ $^{(b)}$  \\
& (J2000) &  (J2000) &  (pc$^2$) & (Jy beam$^{-1}$ km s$^{-1}$) & (km s$^{-1}$) & (km s$^{-1}$) & (Jy km s$^{-1}$) \\ \hline
North         & 1:43:49.2  & -74:32:00.6  & 8.1                & 1.67$\pm$0.05                          & 174.6$\pm$0.6        &  1.33$\pm$0.03    &   30.67$\pm$2.86                         \\
South         & 1:43:53.8  & -74:32:24.7  & 5.8                & 2.81$\pm$0.05                          & 174.0$\pm$0.9       &  1.93$\pm$0.05   &  21.65$\pm2.05$    \\  \hline                       
\end{tabular}
\tablefoot{
\tablefoottext{a}{Peak values obtained from the CO(2$-$1) integrated line image between 172 and 176 km s$^{-1}$.}
\tablefoottext{b}{Values obtained from the gaussian fit to the integrated spectra over the cloud area.}}
\end{table*}

We obtain the integrated velocity spectra of each source, adding all spectra in the source area. The resulting spectra has an rms noise of $\sigma=0.30$ Jy km\,s$^{-1}$ for CO cloud North and $\sigma=0.25$ Jy km\,s$^{-1}$ for CO cloud South, in a velocity channel width 0.5 km\,s$^{-1}$, respectively. We fit a gaussian profile to the velocity integrated spectra of clouds North and South to obtain the central velocity $v_{LSR}$ in km s$^{-1}$, velocity dispersion $\sigma_v$ in km s$^{-1}$ and integrated flux density $S_{CO(2-1)}\Delta v$ in Jy km s$^{-1}$. We adjust the gaussian model to the spectra through least squares fitting.
We correct $\sigma_v$ for broadening due to the finite spectral resolution, using Equation 10 of \citet{rosolowsky2006} to obtain the deconvolved velocity dispersion $\sigma_{v,dc}$, and report the velocity FWHM $\Delta v$ calculated from $\sigma_{v,dc}$. We do not extrapolate $\sigma_v$ to correct for a sensitivity bias (as stated in \citealt{rosolowsky2006}), as the peak emission of the CO clouds North and South is detected with a S/N of 33 and 56, respectively.

The CO cloud North is centered at a velocity of $174.6\pm0.6$ km s$^{-1}$ with a velocity FWHM of $1.33\pm0.03$ km s$^{-1}$, and the CO cloud South is centered at $174.0\pm0.9$ km s$^{-1}$ with a FWHM of $1.93\pm0.05$ km s$^{-1}$. Both clouds are centered at similar velocities and have similar velocity linewidths. The CO North cloud shows a peak $I_{CO(2-1)}$ emission of $1.67\pm0.05$ Jy beam$^{-1}$ km s$^{-1}$ and cloud South has a peak $I_{CO(2-1)}$ emission of $2.81\pm0.05$ Jy beam$^{-1}$ km s$^{-1}$. However, as CO cloud North covers an area of 8.1 pc$^{2}$ while CO cloud South covers 5.8 pc$^{2}$, the North cloud has a larger emission than the South, with a total flux density $S_{CO(2-1)}\Delta v$ of $30.67\pm2.86$ Jy km s$^{-1}$ and $21.65\pm2.05$ Jy km s$^{-1}$, respectively.
 
The results of these fits are summarized in Table \ref{tab:COspectrachar}.

\subsection{Physical properties\label{cloudprops}}

We derive the sizes, CO(2$-$1) luminosities and virial masses for the CO(2$-$1) clouds we found in Magellanic Bridge A using the parameters obtained from the integrated spectra for each cloud.

We calculate the radius
using the second moments of emission, as suggested in Section 2.1 of \citet{rosolowsky2006} using $R=1.91\sigma_{r}D$, where $\sigma_{r}$ is the geometric mean of the spatial second moments of emission in radians, corrected for the spatial broadening due to the beam as indicated by their Equation 9, and $D$ is the distance to the cloud in pc. We assume that Magellanic Bridge A is at a distance of $\sim60$ kpc, same as the distance to the SMC \citep{harries2003}. We do not extrapolate the radius $\sigma_r$ as we do not expect a significant change in the size $\sigma_r$, for the same reason as we did not do it for $\sigma_v$: both CO sources are detected with high S/N ratios as can be seen in Fig. \ref{fig:CO2-1}. The radius obtained for the CO clouds are $1.29\pm0.07$ pc for cloud North and $1.04\pm0.04$ pc for cloud South.

We calculate the CO(2$-$1) luminosities in K km s$^{-1}$ pc$^{2}$ 
using $L_{CO(2-1)} = 611.5 S_{CO(2-1)}\Delta v D^2$, where $S_{CO(2-1)}\Delta$v is the CO(2$-$1) integrated flux density in Jy km s$^{-1}$, obtained from the ALMA and APEX combined CO(2$-$1) line cube, and $D$ is the distance to  Magellanic Bridge in Mpc. The resulting luminosities are 
$L_{CO(2-1)}= 67.5 \pm 6.3$ K km s$^{-1}$ pc$^2$ for CO cloud North, and $L_{CO(2-1)}= 47.7 \pm 4.5$ K km s$^{-1}$ pc$^2$ for CO cloud South. 

We compare the Magellanic Bridge A total luminosity as measured by the single dish APEX CO(2$-$1) observations and by the ALMA CO(2$-$1) observations, to determine any missing diffuse emission. The total luminosity detected by APEX, measured inside a a $58\arcsec \times 58\arcsec$ box centered at $\alpha=$1:43:51.3, $\delta=-$74:32:09.2, is $134 \pm 9$ K km s$^{-1}$ pc$^2$.
The luminosity of clouds North and South in the ALMA CO(2$-$1) line cube
is $66.7\pm6.0$ K\,km\,s$^{-1}$\,pc$^2$ and $48.4\pm4.2$ K\,km\,s$^{-1}$\,pc$^2$, respectively. If we add the CO luminosities of the two clouds, we obtain a total luminosity for Magellanic Bridge A of $115\pm10$ K\,km\,s$^{-1}$\,pc$^2$. Thus, ALMA recovers, within the uncertainties, almost all the luminosity measured in the APEX CO(2-1) line cube, implying that most of the emission in Magellanic Bridge A is concentrated in the two CO(2$-$1) clouds we characterize in this work.

We determine the virial masses $M_{\mathrm{vir}}$ of the clouds using Equation 3 of \citet{maclaren1988}:
\begin{equation}
    M_{\mathrm{vir}} = 190R(\Delta v)^2
\end{equation}
where $R$ is the radius in pc, $\Delta v$ is the FWHM of the line emission in km s$^{-1}$ and $M_{\mathrm{vir}}$ is in M$_{\sun}$. This equation assumes spherical, gravitationally bound molecular clouds and a density profile $\rho \propto r^{-1}$. We use the deconvolved radii in Table \ref{tab:COphysicalprops} and the FWHM in Table \ref{tab:COspectrachar}. The virial mass for CO source North is $437.3\pm26.1$ M$_{\sun}$ and for source South is $732.9\pm37.0$ M$_{\sun}$. Even though these clouds are parsec-sized, they contain $\sim5\times10^2$ M$_{\sun}$, similar to the masses of the most massive molecular clumps in the Milky Way, which have typical radii $\sim1$ pc \citep[see][and references within]{bergin2007}.

We estimate the CO-to-H$_2$ conversion factor $\alpha_{CO}=M_{\mathrm{vir}}/L_{CO}$ and the corresponding $X_{CO}$, which is usually derived for the \element[][12]{CO}(1$-$0) emission. We will not attempt to correct for variations in the $J=2-1$ to $J=1-0$ ratio $r_{21}$, and simply assume $r_{21}\sim1$ for the purposes of comparison, which is also a common result for SMC observations \citep[e.g.,][]{sestpaperIV, sestpaperV}. For CO cloud North, the derived conversion factor is $\alpha_{CO}=6.5\pm0.7$ M$_{\sun}$ (K kms$^{-1}$ pc$^2$)$^{-1}$ ($X_{CO}=(3.0\pm0.3)\times 10^{20}$ cm$^{-2}$ (K km s$^{-1}$)$^{-1}$) and for cloud South, $\alpha_{CO}=15.3\pm1.6$ M$_{\sun}$ (K km s$^{-1}$ pc$^2$)$^{-1}$ ($X_{CO}=(7.1\pm0.7)\times 10^{20}$ cm$^{-2}$ (K km s$^{-1}$)$^{-1}$). These values are 1.5 and 3.6 times the canonical value for the Milky Way $\alpha_{CO}(MW)=4.3$ M$_{\sun}$ (K km s$^{-1}$ pc$^2$)$^{-1}$ ( $X_{CO}=2\times 10^{20}$ cm$^{-2}$ (K km/s)$^{-1}$).

The derived properties $R$, $M_{vir}$, $L_{CO(2-1)}$ and $\alpha_{CO}$ are summarized in Table \ref{tab:COphysicalprops}.

\begin{table*}[ht]
\centering
\caption{\label{tab:COphysicalprops}Physical properties of the emissions found in the $^{12}CO(2-1)$ line cube}
\begin{tabular}{lllll}
\hline\hline
CO Source & $R^{*}$  & $M_{vir}$  & $L_{CO(2-1)}$  & $\alpha_{CO}^{**}$  \\
&(pc) & ($M_{\sun}$) &(K km s$^{-1}$ pc$^2$)& ($M_{\sun}$ (K km s$^{-1}$ pc$^2$) $^{-1}$)
\\ \hline

North         &1.29$\pm$0.07 & 437.3$\pm$26.1          & 67.5$\pm$6.3                        & 6.5$\pm$0.7                                                \\
South         & 1.04$\pm$0.04 & 732.9$\pm$37.0       & 47.7$\pm$4.5                       & 15.3$\pm$1.6  \\ \hline                                          
\end{tabular}
\tablefoot{\tablefoottext{*}{Corrected for spatial broadening.}
\tablefoottext{**}{Assumes a CO(2$-$1) to CO(1$-$0) ratio $r_{21}\sim1$.}}
\end{table*}


\section{Discussion\label{discussion}}

In this section, we discuss our results and compare them to previous observations of Magellanic Bridge A and other low-metallicity galaxies.

\subsection{CO(2-1) in comparison with previous studies}

The molecular clouds we find in CO(2$-$1) emission have similar spectral properties and masses as inferred from previous unresolved CO measurements in Magellanic Bridge A. The central velocities of CO sources North and South, $174.6\pm0.6$ and $174.0\pm0.9$ km s$^{-1}$, respectively, are close to the 174.7 km s$^{-1}$ reported for Magellanic Bridge A by \citet{mizuno2006} through \element[][12]{CO}(1$-$0) emission with NANTEN (beam $FWHM=2.6\arcmin$). The linewidths of both CO clouds, $1.33\pm0.03$ km s$^{-1}$ for cloud North and $1.93\pm0.05$ km s$^{-1}$ cloud South, are consistent with the \element[][12]{CO}(1$-$0) velocity FWHM of 1.6 km s$^{-1}$ reported for Magellanic Bridge A in \citet{mizuno2006} and with the \element[][12]{CO}(3$-$2) FWHM of 1.4 km s$^{-1}$, reported in \citet{muller2014} observed with ASTE Telescope ($FWHM=22\arcsec$). We find that the sum of the virial masses of the CO clouds North and South, $(1.17\pm0.06)\times10^3$ M$_{\sun}$, is in agreement with the mass estimated for Magellanic Bridge A by \citet{mizuno2006}, $10^3$ M$_{\sun}$, using $I_{CO(1-0)}$ and a conversion factor of $X_{CO}=1.4\times10^{21}$ cm$^{-2}$ (K km s$^{-1}$)$^{-1}$.

Comparing the CO luminosities found by previous studies of this source is not easy, as they measure different rotational transitions with different resolutions and sensitivities. Nevertheless, by comparing the luminosities over matched areas, together with the similarities in velocity FWHM and virial mass, we estimate that the emission detected in previous observations is mainly associated to the two clouds we observe in this work. The \element[][12]{CO}(1$-$0) luminosity in the NANTEN detection of Magellanic Bridge A is $L_{CO(1-0)}=70 \pm 8$ K km s$^{-1}$ pc$^{2}$  \citep{mizuno2006}. The \element[][12]{CO}(3$-$2) luminosity in the ASTE detection is $L_{CO(3-2)}=64 \pm 8$ K km s$^{-1}$ pc$^{2}$ \citep{muller2014}. The total CO(2$-$1) luminosity from both Magellanic Bridge A clouds we measure is $L_{CO(2-1)}=115\pm11$ K km s$^{-1}$ pc$^{2}$. The resulting \element[][12]{CO} (3$-$2)/(2$-$1) luminosity ratio is $r_{32}=0.56\pm0.09$, which is reasonable for a star-forming cloud. The \element[][12]{CO} (2$-$1)/(1$-$0) ratio is $r_{21}=1.64\pm0.24$, which is unusually high with respect to typical values observed in the Milky Way, the Magellanic Clouds, and other galaxies where $r_{21}\sim1.1-0.7$ \citep[e.g.,][]{Rubio1996, sorai2001, Bolatto2003, nikolic2007}. Although in localized regions larger ratios can be observed, caused by effective optical depth \citep{Bolatto2003}, it is also possible that the NANTEN \element[][12]{CO}(1$-$0) luminosity is underestimated. Because of the similarities in velocity, FWHM, and luminosity, it seems that most (if not all) of the CO emission from Magellanic Bridge A detected in previous studies comes from the two compact clouds presented in this work.

The CO emission from clouds North and South clouds show interesting differences from parsec-sized CO sources present in the Milky Way. Their virial masses are one order of magnitude higher, implying larger densities in the gas. At radii between 1 and 2 pc, clouds in the Milky Way have masses between a few 10 to a few 100 M$_{\sun}$ \citep{miville2017}, while the clouds found in Magellanic Bridge A have 400 and 700 M$_{\sun}$ for North and South respectively. Note, however, that most of the clouds in the \citet{miville2017} catalog have large virial parameters, and as a consequence are not self-gravitating and are unlikely to be forming stars. 
The assumption of virial, or self-gravitating equilibrium in the Magellanic Bridge A clouds may not be strictly correct, but it is not unreasonable for star-forming clouds. In fact, clouds that have an excess of kinetic energy with respect to their potential energy will be short-lived and very unlikely to locally collapse to form stars. If the CO-emitting cloud is embedded in a large envelope, part of the velocity dispersion may be associated with the external confining pressure associated with the weight of the envelope. In that case the virial mass estimate is, effectively, including part of the mass of the envelope. Using the virial mass estimate, the 
CO clouds North and South have surface densities $\Sigma_{mol}\approx 80$ and 230 M$_\sun$\,pc$^{-2}$ respectively, and corresponding bulk densities $n(H_2)\sim700$ and $2600$ cm$^{-3}$ assuming spherical geometry. The surface densities of Magellanic Bridge A clouds are similar to typical $\Sigma_{mol}$ for star-forming clouds in the Milky Way disk \citep[e.g.,][]{Heiderman2010,Evans2014}, but their bulk volume densities are larger than commonly found bulk densities in our Galaxy as a reflection of the small size of the clouds in this work. \citet{romanduval2010} find $\Sigma_{mol}\sim144$~M$_\sun$\,pc$^{-2}$ and $n(H_2)\sim230$ cm$^{-3}$ as an average for Milky Way clouds in the Galactic Ring Survey, which tend to be star-forming clouds. The Magellanic Bridge clouds have volume densities that are in fact similar to the densities of clumps in local, star-forming molecular clouds 
\citep[e.g.,][]{Evans2014}. This supports the idea that what we see emitting brightly in CO at low metallicities corresponds to the denser regions of molecular clouds. 

In other aspects, the Magellanic Bridge A clouds are similar to clouds found in the Magellanic System and other low-metallicity galaxies. Their CO(2$-$1) emission has similar velocity dispersion and luminosity to other CO clouds studied in the Magellanic Bridge.
Molecular clouds with radii $0.3-1$ pc, velocity FWHM $1-2$ km\,s$^{-1}$ and CO luminosities $10-100$ K\,km\,s$^{-1}$\,pc$^{2}$ have been found in Magellanic Bridge B \citep{saldano2018}. Magellanic Bridge C \element[][12]{CO}(1$-$0) emission is concentrated in clouds with radii $0.9-1.5$ pc and velocity FWHM $0.5-1.4$ km\,s$^{-1}$ \citep{kalari2020}, similar to the clouds we found in this work. The Magellanic Bridge A clouds have similar radii and slightly narrower linewidths and luminosities than regions in the LMC and SMC with active star formation, observed at similar (subparsec) resolution. For example, in 30 Doradus in the LMC, clouds with radii $\gtrsim0.5$ pc show lines $1-6$ km\,s$^{-1}$ wide and CO(2$-$1) luminosities $\gtrsim 100$ K\,km\,s$^{-1}$\,pc$^{2}$ \citep{indebetouw2013}. In N83 in the SMC, clouds with radii $\sim0.8$ pc have linewidths of $\sim4$ km\,s$^{-1}$ and CO(2$-$1) luminosities $100-200$ K\,km\,s$^{-1}$\,pc$^{2}$ \citep{muraoka2017}. The narrower linewidths in Magellanic Bridge A may be the consequence of lower cloud masses, or alternatively lower external pressures \citep{field2011}. Small clouds ($1-3$ pc radii) with linewidths around $2-3$ km\,s$^{-1}$ and high densities ($n(H2) \sim 10^3$ cm$^{-3}$) have been reported in dwarf galaxies like WLM \citep{rubio2015} and NGC6822 \citep{schruba2017}.  
In general, our results support the idea that CO in low-metallicity galaxies traces dense regions of the molecular clouds, and is detected as compact, dense CO clouds with narrow linewidths \citep{rubio2015}.

The $\alpha_{CO}$ CO-to-H$_2$ conversion factor values estimated in this work using the virial masses are around 2-4 times larger than the standard Milky Way disk value $\alpha_{CO}(MW)=4.3$ M$_{\sun}$ (K kms$^{-1}$ pc$^2$)$^{-1}$. Although large conversion factors are observed on large scales for low metallicity galaxies, observations on small scales and in regions where CO is bright in low-metallicity cloud complexes frequently yield conversion factors $\alpha_{CO}$ similar to the canonical value for the Galaxy \citep[e.g.,][]{rubio2015, schruba2017, wong2017, jameson2018}. \citet{rubio1993} found that in the SMC, $X_{CO}$ decreases with cloud size as $\log X_{SMC} = 0.7 \log R + 20.26$ between 10 and 100 pc. Extending their relation for cloud sizes R $\sim1$ pc, the resulting value of $X_{CO}$ is almost the canonical galactic value. For CO cloud North, $\alpha_{CO}$ is in good agreement with the value obtained for clouds sizes $\sim2$ pc using the \citet{rubio1993} relation. Therefore, the  $\alpha_{CO}$ values found in this work are consistent with $\alpha_{CO}$ values found for other low-metallicity regions at similar resolution.

\subsection{Gas and dust comparison \label{sec:gasdustcomp}}

\begin{figure}
    \centering
    \includegraphics[width=\hsize]{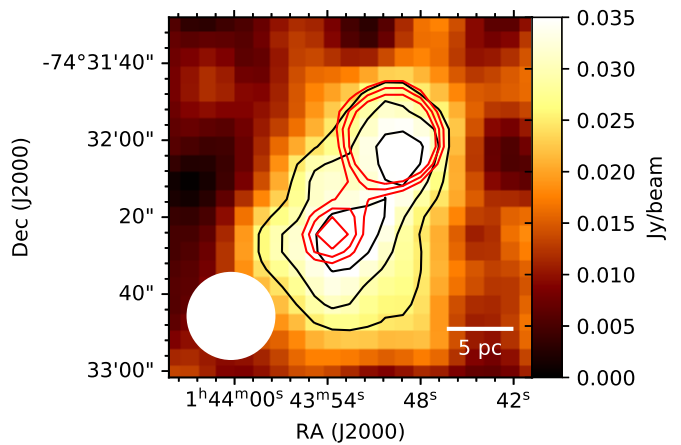}
    \caption{LABOCA continuum image at 870~$\mu$m, with black contours placed at 25, 30 and 35 mJy/beam, as in Figure \ref{fig:contprogression}. The white circle represents the beam size (beam FWHM$=22\arcsec$). Red contours correspond to the ALMA and APEX CO(2$-$1) combined line emission convolved to the APEX resolution of 22\arcsec, integrated between 172 and 176 km s$^{-1}$, at 5$\sigma$, 6$\sigma$ and 7$\sigma$, where $\sigma$ is the rms of the integrated image ($\sigma=1.8$ Jy beam$^{-1}$ km s$^{-1}$). The scalebar at the lower right corner represents a 5 pc length. }
    \label{fig:CO21convolvedlaboca}
\end{figure}

We compare the molecular gas and dust emission in Magellanic Bridge A by convolving the CO(2$-$1) ALMA and APEX combined cube to 22\arcsec\, resolution. We integrate the convolved CO(2$-$1) emission between 172 and 176 km\,s$^{-1}$. The rms noise of the resultant image is $\sigma= 1.8$ Jy beam$^{-1}$ km s$^{-1}$. The CO(2$-$1) peaks can still be distinguished as separate sources at this resolution, but they share the same 3 and 5$\sigma$ contours (see Figure \ref{fig:CO21convolvedlaboca}).

At this common resolution we can characterize the relative sizes of the sources in dust and CO by fitting two-dimensional Gaussians. In CO(2$-$1), the FWHM of the Gaussians are $(7.1\pm0.2)$ pc $\times (6.6\pm0.2)$ pc for cloud North, and $(5.6\pm0.4)$ pc $\times (4.9\pm0.3)$ pc for cloud South. Sources North and South in the 870~$\mu$m continuum are described by Gaussians with FWHM of $(10.3\pm0.6)$ pc $\times (10.0\pm0.7)$ pc and $(16.9\pm1.1)$ pc $\times (16.5\pm0.5)$ pc, respectively. 
The ratio of the continuum to CO spatial size for cloud North source is $1.5\pm0.1$, while for cloud South is $3.0\pm0.3$. 
This difference in size (in particular for cloud South) suggests the presence of an extended halo of cold material, possibly molecular, that is not bright in CO. Indeed, the expectation from photodissociation region models is that at low metallicity the region of a cloud that emits brightly in CO shrinks due to the diminished extinction caused by the lower dust-to-gas ratio \citep{bolatto2013}.

It is interesting to note that not only the size of the emission is larger in the long-wavelength dust continuum, but the gas mass estimates from the FIR dust continuum MBB fit to sources North and South are a factor of $\sim3-4$ larger than virial masses from CO (see Tables \ref{tab:parameters} and \ref{tab:COphysicalprops}). This result is along the lines of previous studies in the SMC, where the total gas mass derived using CO emission is substantially smaller than the cold gas mass obtained from modeling the dust emission \citep{rubio2004,leroy2007,bot2007,bot2010,bolatto2011,jameson2016}. It is consistent with the idea that the CO-emitting regions of molecular clouds at low metallicity are encapsulated in a much larger cold gas envelope, likely molecular, that is not emitting in CO.

\subsection{Gas and dust in the context of star formation}

The North and South clouds in Magellanic Bridge A show differences in the mid-infrarred. In the 8 $\mu$m and 24 $\mu$m emission images of SAGE-SMC program, source North is detected with a flux of $0.9\pm0.1$ mJy and $1.0\pm0.1$ mJy, respectively, whereas source South has a flux density of $7.1\pm0.3$ mJy and $60.0\pm0.7$ mJy, respectively \citep{gordon2011}. The 8 $\mu$m emission is associated with the presence of polycyclic aromatic hydrocarbons (PAHs), which are components of dust and are usually correlated with star formation activity \citep{chen2014}. The 24 $\mu$m emission is a tracer of warm dust, which is associated with massive star formation. The difference in fluxes at 24 $\mu$m is consistent with the presence of a B-type star towards the South cloud. 

The CO(2$-$1) clouds in Magellanic Bridge A have ongoing star formation, with associated YSOs. The North source coincides with a faint YSO candidate (J014349.20-743200.63) classified by \citet{chen2014}, while the South source coincides with a brighter YSO (J014353.94-743224.71) classified as an embedded source in the same work. In fact, the South source hosts a compact multiple system, where a B-type star dominates the far-ultraviolet (FUV) light while the YSO in this location dominates the near-infrarred and mid-infrarred light \citep{chen2014}. The dust temperature reflects this difference in stellar content: the South cloud is being heated by the multiple stars it hosts, in particular a B-type star, while the North cloud might be mostly heated by its fainter YSO. Therefore, it seems that the South molecular cloud has already had an episode of massive star formation, and indeed \citeauthor{chen2014} speculate that the formation of a group of early B stars $5-10$~Myr ago and the corresponding expansion of their \ion{H}{II} regions (still visible in H$\alpha$) may have triggered the current star formation activity in Magellanic Bridge A.


\section{Summary and conclusions\label{conclusions}}

We characterize the molecular gas and dust emission from the Magellanic Bridge A molecular clouds using ALMA 1.3 mm continuum and CO(2$-$1) observations with sub-parsec resolution, together with APEX 870~$\mu$m continuum and CO(2$-$1) line observations at $\sim6$ pc resolution and Spitzer and Herschel FIR archival data.

At this resolution, Magellanic Bridge A separates into two components, North and South. Their dust emission has temperatures of $T_d=21.7\pm2.2$ K and $T_d=24.5\pm1.1$ K respectively, with an emissivity exponent $\beta\approx1.4$ obtained from the MBB fitting to the FIR emission of Magellanic Bridge A at $\lambda\leq500$~$\mu$m. The difference in temperature is consistent with the fact that the North source is not detected at 100 and 160 $\mu$m, and that its star formation activity seems to be weaker than that of the South source. Using the MBB model that best fits the FIR dust fluxes we obtain total gas masses of $(1.3\pm0.3) \times 10^3$ M$_{\sun}$ at source North and $(2.9\pm1.2)\times 10^3$ M$_{\sun}$ at source South, and a total gas mass of $(6.8\pm1.5)\times 10^3$ M$_{\sun}$ for the entire complex. 

After removing possible contributions from free-free and CO line emission, the bolometer measurement at 870~$\mu$m from LABOCA shows a very significant submillimeter excess of a factor of $\sim6.7$ for source North and $3.0$ for source South over their FIR fits. The 870~$\mu$m image exhibits morphology consistent with that of the FIR images, peaking at the location of the CO clouds. The calibration of the data does not appear to be at fault, and the noise in the image is entirely consistent with a priori expectations: in fact, without the excess there would have been a marginal detection of the source. A similar excess, however, is not detected in the ALMA 12m and Morita Array imaging at 1.3 mm. The upper limits for flux density at 1.3 mm are consistent with the predicted fluxes by the MBB model. This requires that either: 1) emission is extended enough that it is filtered out by the interferometer, or 2) that the excess is caused by a process that peaks at 870~$\mu$m \citep[perhaps similar to the Anomalous Microwave Emission detected at lower frequencies,][]{dickinson2018}. High resolution observations at 870~$\mu$m with ALMA will help to establish the nature of the excess.

We find dense molecular clouds in Magellanic Bridge A using ALMA+APEX combined CO(2$-$1) line emission, which spatially coincide with the North and South dust sources. These two clouds have radii $\sim1$ pc, velocity FWHM of 1.3 and 1.9  km s$^{-1}$ and virial masses of $\sim400$ and 700 M$_{\sun}$ for clouds North and South, respectively. Accordingly, their bulk volume densities are $n(H_2)\sim 700$ and $2600$~cm$^{-3}$, significantly higher than the typical density of clouds in the Milky Way \citep[$n\sim230$~cm$^{-3}$,][]{romanduval2010}. The total interferometric $L_{CO(2-1)}$ of the complex is 85\% of the luminosity $L_{CO(2-1)}$ measured with the APEX single-dish, which suggests that most (if not all) CO(2$-$1) emission from Magellanic Bridge A comes from these two clouds. Using the virial mass, we find a CO-to-H$_2$ conversion factor $\alpha_{CO}$ of $6.5\pm1.2$ M$_{\sun}$ (K kms$^{-1}$ pc$^2$)$^{-1}$ and $15.3\pm1.6$ M$_{\sun}$ (K km s$^{-1}$ pc$^2$)$^{-1}$ for the North and South clouds respectively, consistent with the trend of $\alpha_{CO}$ approaching Milky Way values on the small spatial scales.
 
We compare the CO(2$-$1) and dust emissions at a common resolution of 22\arcsec. We find that CO(2$-$1) emission covers a smaller area than continuum emission in the South cloud, whereas they have a similar spatial extent on the more quiescent North cloud. The total gas mass derived from the dust emission $\sim 4$ times larger than the sum of virial masses obtained for both clouds.


\begin{acknowledgements}
We thank the anonymous referee for very constructive feedback and comments which helped to improve the quality of this article. M. T. V. acknowledges financial support from CONICYT (Chile) through the scholarship CONICYT-PFCHA/
Magister Nacional 2018 - 22180279, Universidad de Chile VID grant ENL22/18 and FONDECYT grant No1190684. M.R. wishes to acknowledge support from Universidad de Chile VID grant ENL22/18, from CONICYT (CHILE) through FONDECYT grant No1190684, and partial support from CONICYT project Basal AFB-170002.  H.P.S acknowledges financial support from a fellowship from Consejo Nacional de Investigaciones Científicas y Técnicas (CONICET), and from Secretaría de Ciencias y Técnicas (Secyt), Córdoba, Argentina. This paper makes use of the following ALMA data: ADS/JAO.ALMA\#2012.1.00683.S. ALMA is a partnership of ESO (representing its member states), NSF (USA) and NINS (Japan), together with NRC (Canada), MOST and ASIAA (Taiwan), and KASI (Republic of Korea), in cooperation with the Republic of Chile. The Joint ALMA Observatory is operated by ESO, AUI/NRAO and NAOJ. This publication is based on data acquired with the Atacama Pathfinder Experiment (APEX). This research made use of APLpy, an open-source plotting package for Python (Robitaille and Bressert, 2012). This research made use of Astropy,\footnote{http://www.astropy.org} a community-developed core Python package for Astronomy \citep{astropy:2013, astropy:2018}.
\end{acknowledgements}

%
%
\bibliographystyle{aa}
\bibliography{main}

\end{document}